\documentclass[twocolumn,
preprintnumbers,
amsmath,
amssymb,
aps,
prl,
superscriptaddress
]{revtex4}
\usepackage{graphicx}
\usepackage{textcomp} 
\usepackage[caption=false]{subfig} 
\begin{document}
\title{\textbf{Self-Similarity and Energy Dissipation in Stepped Polymer Films}}
\author{Joshua D. McGraw}
\address{Department of Physics \& Astronomy and the Brockhouse Institute for Materials Research, McMaster University, Hamilton, Canada}
\author{Thomas Salez}
\address{Laboratoire de Physico-Chimie Th\'eorique, UMR CNRS Gulliver 7083, ESPCI, Paris, France}
\author{Oliver B\"{a}umchen}
\address{Department of Physics \& Astronomy and the Brockhouse Institute for Materials Research, McMaster University, Hamilton, Canada}
\author{Elie Rapha\"{e}l}
\address{Laboratoire de Physico-Chimie Th\'eorique, UMR CNRS Gulliver 7083, ESPCI, Paris, France}
\author{Kari Dalnoki-Veress}\email{dalnoki@mcmaster.ca}
\address{Department of Physics \& Astronomy and the Brockhouse Institute for Materials Research, McMaster University, Hamilton, Canada}
\date{\today}

\begin{abstract}
The surface of a thin liquid film with nonconstant curvature is unstable, as the Laplace pressure drives a flow mediated by viscosity. We present the results of experiments on one of the simplest variable curvature surfaces: a stepped polymer film. Height profiles are measured as a function of time for a variety of molecular weights. The evolution of the profiles is shown to be self-similar. This self-similarity offers a precise measurement of the capillary velocity by comparison with numerical solutions of the thin film equation. We also derive a master expression for the time dependence of the excess free energy as a function of the material properties and film geometry. The experiment and theory are in excellent agreement and indicate the effectiveness of stepped polymer films to elucidate nanoscale rheological properties. 
\end{abstract}
\maketitle

The properties of polymers and indeed all molecules in thin films and at interfaces continue to stimulate debate. The apparent increased mobility of polymers at interfaces in both the liquid~\cite{shin07NMT, bodiguel06PRL} and glassy states~\cite{fakhraai08SCI}, their structural properties~\cite{si05PRL, baumchen09PRL}, as well as the effects of preparation and treatment history~\cite{barbero09PRL, raegen10PRL, thomas11PRE, ranxing12MAC} are areas of concerted effort. While length scales of these systems are frequently just a few molecular diameters, the physics governing their liquid state evolution is quite general and can be used to model, for example, geophysical flows~\cite{refBcite}.

When a flat liquid film is in contact with a substrate of lower surface energy, a hole can nucleate to expose some of the underlying surface. This phenomenon induces a nonconstant curvature of the film driven by interfacial tension and leads to a decrease of the total free energy as the hole grows. This process is called dewetting~\cite{brochard90CJP, reiter92PRL}. There have been many studies concerned with details of the shape and evolution of the rim which collects the dewetted fluid~\cite{seemann01PRL, munch06JEM, vilmin06EPJE, fetzer07LAN, baumchen09PRL, reiter09EPJJ, snoeijer10PRE, baumchen2012JCP}, and much has been learned about and from the process of turning a flat film into a collection of droplets. 

Several techniques that have been used~\cite{kerle01MAC, buck04MAC, jdmlev1, teisseire11APL, zhu11PRL, rognin11PRE, rognin12JVS} to provide insight into the aforementioned problems~\cite{shin07NMT, bodiguel06PRL, fakhraai08SCI, si05PRL, baumchen09PRL, barbero09PRL, raegen10PRL, thomas11PRE, ranxing12MAC} take the opposite approach. They rely on the property that curved interfaces have more surface area than flat ones. Such curved surfaces can be driven to flatten by the surface tension, $\gamma$, but are mediated by the viscosity, $\eta$. In several of the surface tension-driven flow techniques, the nanoscale topography is imprinted using a square wave pattern that varies in one lateral dimension only~\cite{teisseire11APL, zhu11PRL, rognin11PRE, rognin12JVS}. In these studies, the authors typically consider the evolution of the amplitude of the perturbation, rather than that of the profile shape. As in many of the dewetting studies, this Letter is concerned with the details of a capillary-driven flow profile, but in a different geometry: a \emph{stepped film}. 

 \begin{figure}[b!]
  \begin{center}     
     \includegraphics{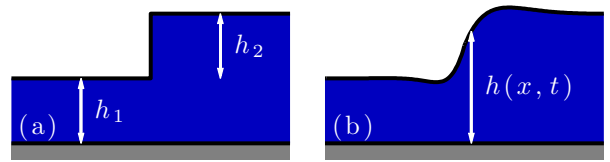}  
  \end{center}
\caption{(a) Schematic of the as-prepared samples. (b) After annealing above the glass transition temperature, the region of transition between the two terrace heights broadens, and a flow profile can be measured. }
  \label{schem}
\end{figure} 

\begin{figure*}[t!]
  \centering
  \subfloat{\includegraphics{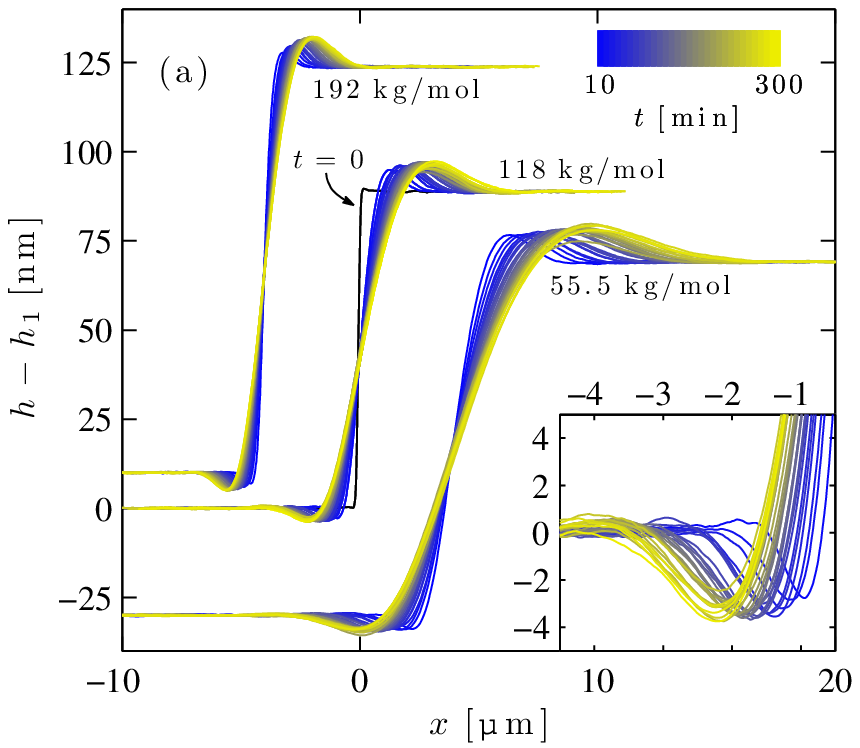}}
  ~ 
  \subfloat{\includegraphics{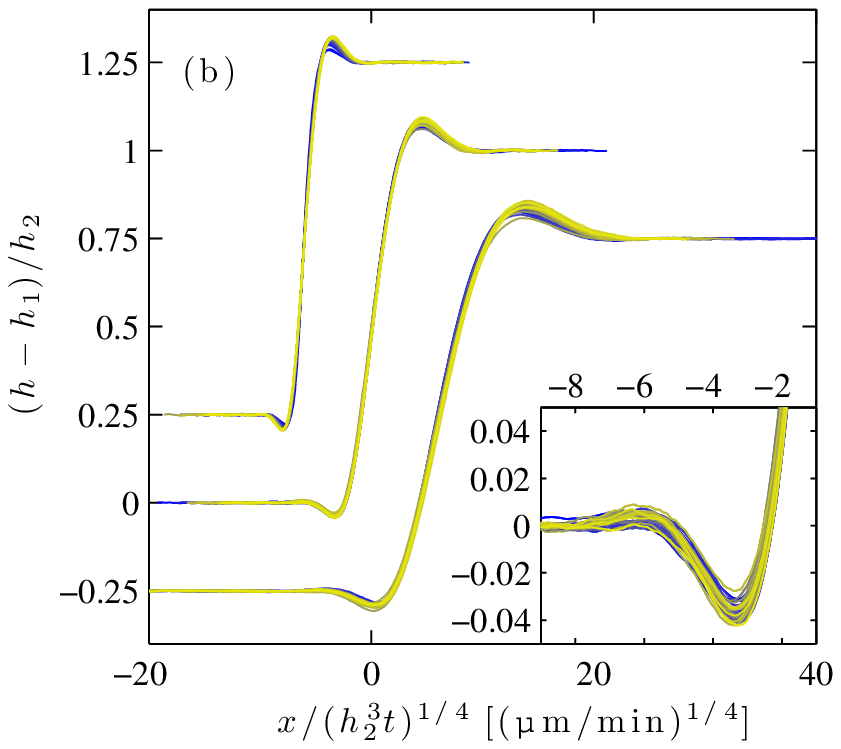}}
  ~ 
  \caption{(a) Height as a function of position and time ($10 \leq t \leq 300$~min) for three stepped films with molecular weights as indicated. From left to right, $\{h_1, h_2\} = \{106, 114\}, \{89, 89\}$, and $\{101, 99\}$~nm.  The inset shows a detail of the `dip' region for the 118 kg/mol sample. (b) Scaled height as a function of scaled position,  according to Eq.~(\ref{adim}); the inset shows the same region as the inset of panel (a) in scaled variables. Whereas the volume constraint uniquely determines the origin, the 192 and 55.5~kg/mol data have been shifted horizontally and vertically in both (a) and (b) for clarity.}
\label{tiger}
\end{figure*}

In a previous contribution~\cite{jdmlev1}, we explained how to prepare the simple geometry of stepped polymer films. The samples are prepared as schematically shown in Fig.~\ref{schem}(a). They are comparable to the square wave patterns discussed above~\cite{teisseire11APL, zhu11PRL, rognin11PRE, rognin12JVS} in that the height is a function of only one spatial dimension and time, $h=h(x,t)$. They are different in that they have only one height step. This two-dimensional geometry allows us to study the broadening in the region over which the height changes from $h_1$ to $h_1+h_2$ (see Fig.~\ref{schem})  in isolation, rather than having to consider the flow from neighboring steps in a periodic geometry. With the use of scaling laws, it was shown that by measuring a particular lateral length scale, one could measure the relative viscosity~\cite{jdmlev1}. In this Letter, we provide a theoretical treatment of the governing thin film equation which is in agreement with the measured height profiles over a wide range of thickness combinations and over several orders of magnitude in viscosities. We explain in detail the dependencies of the energy dissipation on the film geometry and capillary velocity, $\gamma/\eta$. 

Films were prepared by spin-coating polystyrene (PS) dissolved in toluene onto two types of substrates: $1\times1\textrm{ cm}^2$ Si wafers (University Wafer) rinsed with ultrapure water (18.2 M$\Omega\textrm{\,cm}$, Pall, Cascada LS), methanol, and toluene (Fisher Scientific, Optima grade), as well as freshly cleaved mica substrates (Ted Pella, Inc.). The PS molecular weights were $M_w = 15.5,$ 55.5, 118 and 192~kg/mol with polydispersity indices $\leq1.07$ (Polymer Source, Inc). The film thicknesses ranged from 30 to 200 nm. These heights were always much larger than the typical size of molecules making up the films. The as-cast samples were annealed at $130\ ^\circ\textrm{C}$ in a vacuum oven ($10^{-5}\textrm{ mbar}$) for 24 hours (more than two orders of magnitude greater than the longest relaxation time of the largest polymer used~\cite{bach03MAC}).

The preparation of stepped films proceeded as detailed previously~\cite{jdmlev1}. Briefly, PS films with thickness $h_2$ on a mica substrate were floated onto ultra pure water and transferred to a Si wafer. After drying, these wafers were split along a crystal axis. The split films were again floated onto water and picked up using a Si wafer coated with a PS film of height $h_1$, thus creating a bilayer with step size $h_2$ (see Fig.~\ref{schem}). Samples are described as having geometry $\{h_1, h_2\}$. All sections of film observed in this study had straight edges as viewed with optical and atomic force microscopy (AFM, Veeco Caliber) for distances of a minimum of 50~\textmu m. Figure~\ref{tiger}(a) at $t=0$ illustrates an example of the initial stepped profile; see also Ref.~\cite{jdmlev1}.

Prior to measurement of each profile evolution, $h_2$ was measured (as with all heights reported) using AFM. Then, the film was annealed for $10\textrm{ min at }140\ ^\circ\textrm{C}$ on a hot stage (Linkam) using a heating rate of $90\ ^\circ\textrm{C/min}$ in air~\cite{footnote}. After cooling to room temperature ($\sim 40\ ^\circ\textrm{C/min}$ to below the glass transition, $T_{\mathrm{g}}\sim 100\ ^\circ\textrm{C}$), the step profile was measured. Given $h(x,t)$ as the distance between the substrate-polymer and air-polymer interfaces, $h(x,t)-h_1$ is the quantity measured. For the 55.5, 118 and 192 kg/mol samples, additional annealing for 10 min periods was performed, with the profile evolution obtained at room temperature. After a final measurement of the height profile, a scratch in the PS film was made down to the substrate to measure $h_1$.

Figure~\ref{tiger}(a) shows the time evolution ($10 \leq t \leq 300\textrm{ min}$) of three stepped  polymer films with $h_1 \approx h_2 \approx 100$~nm, each with a different molecular weight. On each stepped film, there is a prominent `dip' on the thin side of the film and a `bump' on the thick side, which is consistent with our earlier observations~\cite{jdmlev1} and can also be seen in the long wave limit of the data presented by Rognin \emph{et al.}~\cite{rognin11PRE}. The evolution proceeds such that positions of the profile extrema are separated over wider distances: the film flattens with time. 

The flow profile can be understood from the Laplace pressure, which arises due to curvature at the fluid interface~\cite{degennes03TXT}; the range of heights considered here allows us to neglect gravitational~\cite{refBcite} and disjoining~\cite{seeman2001PRL} pressures. Assuming that the height gradients are small, the lubrication approximation applies, and the local Laplace pressure is given by $p(x,t) \approx -\gamma \partial_x^{\,2}h$. Though the as-prepared stepped films have steep gradients, they also contain large curvature gradients in the same regions. These high-gradient regions then flow quickly compared to the experimental annealing times and the lubrication approximation is valid for all experiments presented here. As a stepped film levels, the bump represents a region of high pressure relative to the flat regions on the thick side of the film, while the dip has a low pressure relative to flat regions on the thin side. Thus, there are pressure gradients along the region of transition between $h_1$ and $h_1+h_2$ and the flow continues.

To understand the data of Fig.~\ref{tiger}(a) quantitatively, we combine the lubrication approximation and the Stokes equation of viscous flows that leads to
$\partial_xp = \eta \partial_z^{\,2} v$, 
where $v(x,z,t)$ is the horizontal velocity and $z$ is the vertical coordinate~\cite{stillwagon88JAP, oron97RMP, craster09RMP}. Since the longest relaxation time of any polymer used here is of order 100~s, and all leveling times are significantly longer than this time, viscoelastic effects can safely be neglected. Assuming no slip at the solid-liquid interface and no stress at the liquid-air interface gives a Poiseuille velocity profile along $z$. Invoking conservation of volume and using the Laplace pressure  introduced above leads to the thin film equation

\begin{equation}
\partial_th + \frac{\gamma}{3\eta}\partial_x\left(h^3\partial_x^{\,3}h\right) = 0\ .
\label{hxt}
\end{equation}

This differential equation can be nondimensionalized by letting $H=h/h_2$, $X=x/x_0$ and $T=\gamma h_2^{\,3} t/ 3\eta x_0^{\,4}$, where $x_0$ is a typical horizontal length scale of the problem~\cite{stillwagon88JAP}. Furthermore, introducing the variable

\begin{equation}
U=\frac{X}{T^{1/4}} = \left(\frac{3\eta}{\gamma}\right) ^{1/4}\frac{x}{(h_2^{\,3}t)^{1/4}} \ ,
\label{adim}
\end{equation}

one can show that self-similar solutions of the first kind of the form $H(X,T) = F(U)$ exist~\cite{aradian01EPL, Barenblatt1996, refBcite, bowen}. Using a numerical scheme~\cite{bertozziAMS98, salez12a}, we have shown that this self-similarity is satisfied for times $T\gtrsim10^{-4}$ for a profile with $h_1 = h_2 $. We have furthermore verified that the self-similarity is not sensitive to the initial height profile, provided that the heights at $U = \pm\infty$ are constant and different. Therefore, if the interface profile is self-similar, a rescaling of the horizontal axis must collapse all of the experimental data. In Fig.~\ref{tiger}(b) we show the same data as in Fig.~\ref{tiger}(a) but plotted as a function of $x/(h_2^{\,3}t)^{1/4}$. Having done so, we see that the data for all three molecular weights do collapse onto three individual curves, thus demonstrating the self-similarity of the evolution through time. Examining the scaled curve for the 192 kg/mol PS in Fig.~\ref{tiger}(b) closely, one can see that the early time data do not collapse perfectly onto one single curve. This result is due to the fact that it takes longer times to reach the self-similar solutions to Eq.~(\ref{hxt}) for higher molecular weights, at a given temperature. 

In the previous rescaling, the horizontal length scale characterizing the transition between undisturbed film heights is different for each of the three data sets. Since the samples all started with the same initial condition, they have different leveling speeds because the viscosity varies with $M_w$.

\begin{figure}[b!]
  \begin{center}
     \includegraphics{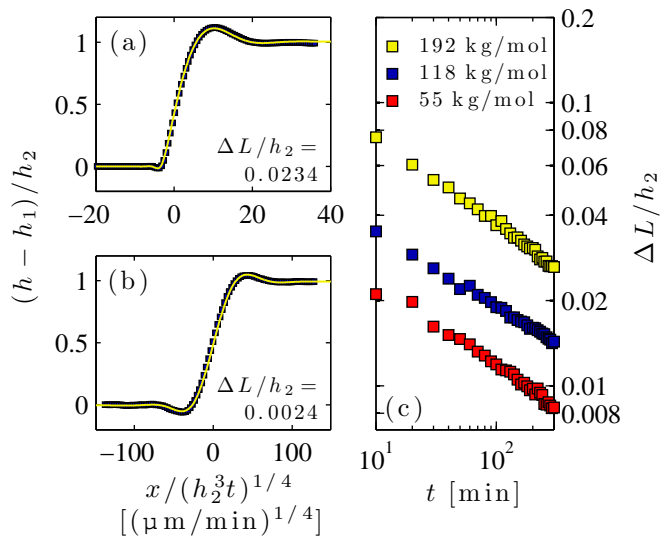}  
  \end{center}
\caption{(a) Rescaled measured profile for a 15~kg/mol PS stepped film with $\{h_1, h_2\} = \{30,193\}$~nm after 10 min of annealing at $140\ ^\circ$C; (b) an identically annealed stepped film with $\{h_1, h_2\} = \{174, 32\}$~nm. The lines in (a) and (b) are horizontally stretched numerical solutions from Eq.~(\ref{hxt}) using Refs.~\cite{bertozziAMS98,salez12a}. (c) Relative excess contour length of the data in Fig.~\ref{tiger} as a function of time; see Eq.~(\ref{eqener}).}
  \label{time}
\end{figure} 

To obtain precise measurements of these leveling speeds, we numerically solve the dimensionless form of Eq.~(\ref{hxt}) with a stepped initial condition of aspect ratio $r=h_1/h_2$. We then fit the computed self-similar profile, $F_{r}(U)$, to the corresponding rescaled experimental profile. Figures~\ref{time}(a) and \ref{time}(b) show the rescaled measured profiles for two 15 kg/mol PS stepped films with different $\{h_1, h_2\}$ values, as well as the corresponding fits to the numerical solutions of Eq.~(\ref{hxt})~\cite{salez12a}. Despite the initial transient flow not described by the lubrication approximation, all experimental profiles presented here approach the self-similar solutions obtained from Eq.~(\ref{hxt}). 
In the fitting procedure, the only free parameter is a horizontal stretch of the computed profile given by the factor $(\gamma/3\eta)^{1/4}$ according to Eq.~(\ref{adim}). This procedure thus yields a measured value of $\eta/\gamma$. Using $\gamma \approx 30~\textrm{mJ/m}^2$~\cite{wuPpoly} for all polymers here, we get $\eta = 7.3\times10^3$, $1.4\times10^5$, $1.1\times10^6$, and $9.1\times10^6$~Pa\,s for $M_w = 15.5$, 55.5, 118, and 192~kg/mol PS at $140\ ^\circ\textrm{C}$. These viscosities follow the expected molecular weight dependence of the polymer melt viscosity~\cite{rubin03TXT} and are in agreement with viscosities measured in the bulk~\cite{bach03MAC}. We have measured $\eta/\gamma$ for nine additional 15.5 kg/mol PS stepped films all with different $\{h_1, h_2\}$ values. We find that the value obtained is independent of $\{h_1, h_2\}$, and the ratio of the standard deviation to the mean is 0.12. 

\begin{figure}[t!]
  \begin{center}
     \includegraphics{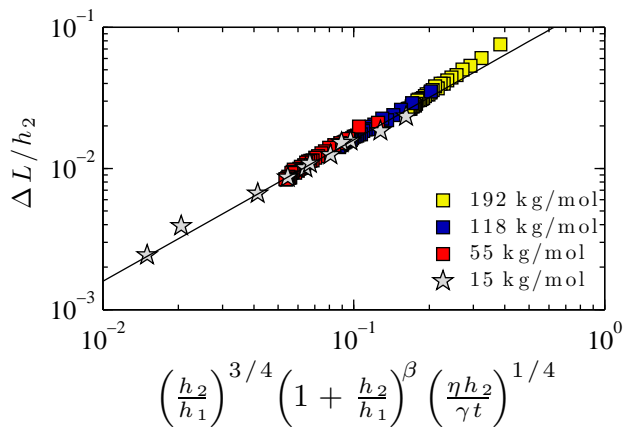}  
  \end{center}
\caption{Relative excess contour length as a function of time, the geometrical and physical parameters of the stepped films, and having made use of $\beta = -0.42$. Data points represent 15 kg/mol films at $t=10\ \textrm{min}$ with different $\{h_1, h_2\}$ values ($0.16 < r < 5.6$), as well as the temporal series of Fig.~\ref{time}(c). The solid line is the theoretical prediction of Eq.~(\ref{final}) with no free parameter ($\xi = 0.16$).} 
  \label{geometry}
\end{figure} 

Having exhibited the self-similarity of the profiles and having set up a robust method to measure the capillary velocity, we now turn to the study of the dissipation law in stepped films. The excess free energy per unit length along the dimension of invariance, $\Delta\mathcal{F}$, is dominated by the interfacial contribution. For small slopes 
\begin{equation}
\label{eqener}
\Delta\mathcal{F}=\gamma\Delta L \approx \frac{\gamma}{2} \int dx\ (\partial_xh)^2\ ,
\end{equation}
where $\Delta L$ is the excess contour length of the profile with respect to the flat limit at infinite time. In Fig.~\ref{time}(c), we plot the relative excess contour length as a function of time for the three samples of Fig.~\ref{tiger}. It is evident that the individual data sets obey a power law in time. Referring also to the two stepped films shown in Figs.~\ref{time}(a) and \ref{time}(b), we see that the contour lengths of these two profiles are not identical despite having been annealed under the same conditions. Therefore, there are temporal and geometrical dependencies on the energy dissipation.

In the following, we develop a theoretical model in order to understand the results of Fig.~\ref{time}. Making the quantities in the integrand of Eq.~(\ref{eqener}) dimensionless as presented above and invoking the self-similarity of the profile, we find that the relative excess contour length remaining after time $t$ satisfies 

\begin{subequations}\label{energy}
\begin{align}
\Delta L&= C(r) \left(\frac{\eta h_2^{\,5}}{\gamma t}\right)^{1/4} \ , \label{etsc}\\ 
\mathrm{with} \ C(r) &= \frac{3^{1/4}}{2}\int dU\ [F_r'(U)]^2\ ,\label{bilc}
\end{align}
\end{subequations}

where $C(r)$ is determined by the sample geometry through the dimensionless self-similar profile $F_r(U)$.

The stepped film problem contains three relevant combinations of heights: $h_2$ for the step height which provides the typical driving force for leveling, as well as $h_1$ and $h_1+h_2$ for the thicknesses of the thin and thick regions of the film. Thus, having already extracted the $h_2$ dependence in Eq.~(\ref{etsc}), the simplest ansatz for a height dependence of $C(r)$ is a power law of the form  
\begin{equation}
\label{eqc}
 C(r) = C(1)\  r^\alpha\left(\frac{1+r}{2}\right)^\beta \ . 
 \end{equation}
This scaling expression connects the case of arbitrary aspect ratio, $r$, to the $r=1$ geometry; normalization to $r = 1$ is ensured by the factor of 2. In order to determine the two exponents, we consider the limit $r\gg1$, where the step is a small perturbation of the flat film. In this limit, Eq.~(\ref{hxt}) can be linearized and solved analytically and we find $C(r)\approx\xi\ r^{-3/4}$, where $\xi\approx 0.16$ \cite{Salez12b}. By comparison to Eq.~(\ref{eqc}) when $r\gg1$, we get $\alpha+\beta = -3/4$ and $\beta = \log[C(1)/\xi]/\log(2)$. Then, using the numerical solutions of Eq.~(\ref{hxt}) presented above, we calculate $C(1)\approx0.12$ and thus $\beta\approx-0.42$.
Finally, we obtain a master expression for the relative excess contour length

\begin{equation}
\frac{\Delta L}{h_2} =\xi\ \left(\frac{h_2}{h_1}\right)^{3/4}\left(1+\frac{h_2}{h_1}\right)^{\beta}\left(\frac{\eta h_2}{\gamma t}\right)^{1/4}\ . \label{final}
\end{equation}

Figure~\ref{geometry} shows the relative excess contour length for all samples and times presented in this study as a function of the combination of $\{h_1, h_2\}, \gamma/\eta$ and $t$ suggested by Eq.~(\ref{final}). The solid line corresponds to the theoretical prediction of Eq.~(\ref{final}) with \emph{no free parameter}. The agreement between the data and Eq.~(\ref{final}) validates the dissipation law, as well as the general scaling to the $r=1$ symmetric case, within the range of our experimental data. As discussed above, the dissipation law is modified for $t<100$~min for the 192~kg/mol PS, where self-similarity may not be valid. Although we have not accessed these limits, we expect deviations when $r \rightarrow 0$ for which the dependency on $h_1$ may differ, and when $t \rightarrow 0$ for which the lubrication approximation is not valid.

In conclusion, we have demonstrated that the thin film equation captures the time evolution of stepped polymer films for a wide range of aspect ratios and material properties. With the use of a numerical solution of this equation, it is now straightforward to precisely measure the capillary velocity of any nonvolatile fluid prepared as described. We further demonstrated the self-similarity of the evolution of the profiles. Finally, details of how the excess surface energy is dissipated by viscosity over time have been obtained. We have determined that the surface energy decreases with a $-1/4$ power law in time. The rate at which energy is dissipated depends on the capillary velocity and on the heights of the initial stepped film. In particular, we have shown through a master expression that the evolution of any stepped film can be rescaled to the $h_1=h_2$ symmetric step. 
\newpage

The authors thank NSERC of Canada,  the \'Ecole Normale Sup\'{e}rieure of Paris, the German Research Foundation (DFG) under Grant No. BA3406/2, the Chaire Total-ESPCI, and the Saint Gobain Fellowship for financial support.


\begin{thebibliography}{38}
\bibitem{footnote} We have verified that the leveling results are unchanged when annealing in a dry nitrogen atmosphere.
\end{thebibliography}
\end{document}